\documentclass[aps, prl, twocolumn,amsmath,amssymb,floatfix,superscriptaddress,prl]{revtex4-2}
\usepackage{leading}
\usepackage[latin3]{inputenc}
\usepackage[makeroom]{cancel}
\usepackage{graphicx}
\usepackage{amsmath}
\usepackage{amsfonts}
\usepackage{chngcntr}
\usepackage{amssymb}
\usepackage{color}
\usepackage[left=2cm,right=2cm,top=2cm,bottom=2cm]{geometry}
\usepackage{graphicx} 
\usepackage{dcolumn}
\usepackage{bm}
\usepackage{simplewick}
\usepackage{array}
\usepackage{appendix}
\usepackage{cancel}
\usepackage[normalem]{ulem} 
\RequirePackage[
   hyperindex,colorlinks,bookmarksnumbered,
   plainpages=true,pdfstartview=FitH]{hyperref}
\hypersetup{linkcolor=blue,urlcolor=blue,citecolor=blue}
\usepackage{hyperref}
\usepackage{mathrsfs}
\usepackage{ragged2e}
\newcommand{\pdag}{{\phantom{\dagger}}}
 \renewcommand{\emph}[1]{\textit{#1}}

\definecolor{darkblue}{rgb}{0,0,0.5}
\definecolor{darkgreen}{rgb}{0,0.5,0}
\definecolor{darkred}{rgb}{.7,0,0}
\definecolor{purple}{rgb}{0.5,0,0.6}
\definecolor{orange}{rgb}{1,0.5,0}
\definecolor{grey}{rgb}{.6,.6,.6}
\definecolor{lightpink}{rgb}{1,0.7,0.75}
\definecolor{pink}{rgb}{1,0.4,0.58}
\definecolor{deeppink}{rgb}{1,0.08,0.58}
\usepackage{upgreek}

\begin{document}


\title{Emergent parafermions with tunable localization in a quantum circuit network}


\author{D. B. Karki}
\email{deep.b.karki@gmail.com}
\affiliation{The Abdus Salam International Centre for Theoretical Physics (ICTP), Strada Costiera 11, I-34151 Trieste, Italy}
\author{Andrew K. Mitchell}
\email{Andrew.Mitchell@ucd.ie}
\affiliation{School of Physics, University College Dublin, Belfield, Dublin 4, Ireland}
\affiliation{Rinn Quantum and Centre for Quantum Engineering, Science, and Technology,\\School of Physics, University College Dublin, Dublin 4, Ireland}


\begin{abstract}
\noindent Quantum circuit simulators with hybrid metal-semiconductor dots provide a versatile platform to realize Kondo anyons. These arise at critical points driven by frustrated Kondo screening of dot charge degrees of freedom and have been demonstrated experimentally in linear circuits of one or two dots. Here we study a triangular triple dot device, where geometric frustration gives rise to an emergent $\mathbb{Z}_3$ parafermion shared among the three dots. We show that the parafermion can be tunably relocated to any pair of dots by applying gate voltages, and is supported in both the weak-tunneling `Kondo limit' and the strong transmission `quasiballistic' regime. The critical states survive strong coupling to external leads and other dots, 
establishing the basis for a scalable and reconfigurable network architecture in which non-Abelian excitations can be realized and manipulated.
\end{abstract}
\maketitle


\textbf{\emph{Introduction.--}} 
Nanoelectronic devices involving quantum dots provide a versatile platform to study strongly correlated electron physics~\cite{GP_Review_2005,meirav1990single,Goldhaber_nat,Cronenwett_1998,koven,keller2014emergent}. Circuits with small spin-based semiconductor dots can be engineered and tuned \textit{in situ} to realize exotic states of quantum matter near impurity quantum critical points~\cite{Vojta2006ImpurityTransitions}, driven by frustration when multiple Kondo screening channels compete~\cite{Potok_NAT,Keller_2015,Mitchell_2012a,*Sela_2011,sen2024many,mitchell2021so,*liberman2021so,Mitchell_2012b}. In particular, the two-channel Kondo (2CK) effect supports intrinsically fractionalized excitations at low energies~\cite{Wiegmann_1983,ad,al1}, with an emergent Majorana degree of freedom~\cite{EK1, Sela_2018,kam}. The interplay between geometric frustration and Kondo physics has also been studied in semiconductor triple-dot setups~\cite{mitchell2009quantum,*mitchell2013local,numata2009kondo,mitchell2010two}. Indeed, circuit networks of coupled dots constitute a platform for quantum simulation of certain Hubbard-type lattice models~\cite{Barthelemy2013QuantumSimulation,Hensgens2017Hubbard,Dehollain2020Nagaoka,VanDiepen2021Heisenberg}. 

More recently, a new quantum circuit paradigm has emerged, based on a hybrid metal-semiconductor architecture~\cite{Pierre_2015,Pierre_2018,han2022fractional,entt,piquard2023observing}. Here, quantum Hall contacts provide unprecedented control and tunability of couplings, while the nominally-identical metallic islands offer a scalable route to simulating novel forms of synthetic quantum matter in uniform arrays or networks~\cite{ccd,Sriram2026}.

The physics of such circuits is described by Matveev's charge-Kondo scenario~\cite{Matveev_1991, Matveev1995, Furusaki1995b}. Each charge-Kondo site consists of a metallic grain with effectively vanishing level spacing $\Delta$ but finite charging energy $E_c$. Spin-polarized electrons tunnel on or off the island from leads connected at quantum point contacts (QPCs) with transmission $\tau$, and the island's charge $N_g$ is adjustable with a gate voltage. Strong correlation physics emerges at low temperatures $\Delta \ll T\ll E_c$. The lowest-energy pair of island charge states forms a pseudospin that is flipped by electronic tunneling at attached QPCs, giving an effective quantum impurity model when the QPC coupling is weak. For one channel~\cite{piquard2023observing} this captures emergent Kondo physics, where the island charge pseudospin is screened by conduction electrons through a non-perturbative process at energies below the Kondo scale $T_K$. Intriguingly, the same low-energy physics is obtained in the opposite limit where the QPCs are nearly perfectly transmitting and a macroscopic number of charge states participate~\cite{Matveev_1991, Matveev1995, Furusaki1995b,Sela_2016,top1}. 

When multiple leads are attached, multichannel Kondo critical points can be probed~\cite{Pierre_2015,Pierre_2018,han2022fractional,entt}. This provides a possible route to engineering one or more Kondo anyons~\cite{Sela_2018,lopes2020anyons,kam} in coupled island circuits. The two-site device design of Ref.~\cite{ccd} allows to couple two critical 2CK Majorana zero modes, as explored recently in Ref.~\cite{karki2026coupled}. However, at the charging triple-point of the experiment~\cite{ccd}, a more exotic $\mathbb{Z}_3$ parafermion state could also be found~\cite{KBM1,KBM}. Series-coupled charge-Kondo chains have since attracted considerable attention~\cite{dde1,dpk1,viet, Karki_2026,lm1,es1}, especially for their anomalous charge and heat transport properties~\cite{amm,Slobodeniuk_2013,kkr,genr,eu1,para,oliever,es1}.

\begin{figure}[b]
    \centering
    \begin{minipage}[t]{0.652\linewidth}
        \vspace{0pt}
        \includegraphics[width=\linewidth]{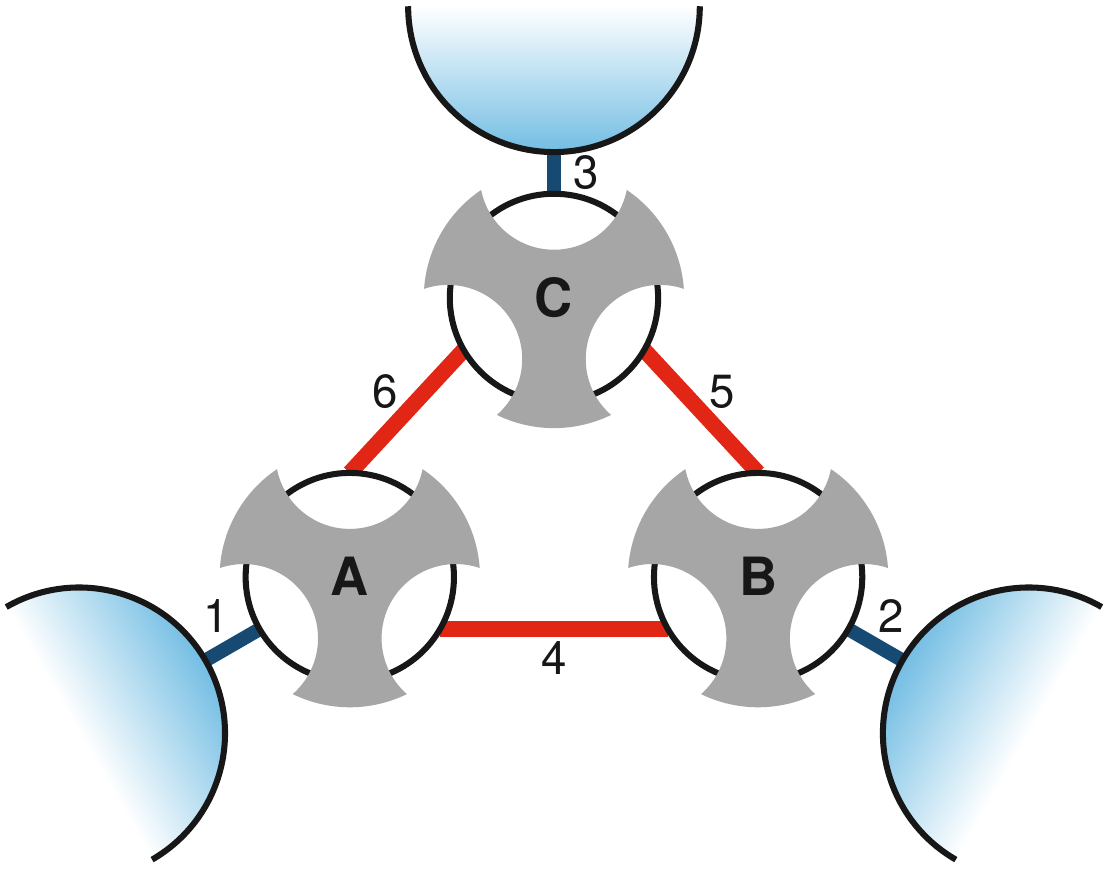}
    \end{minipage}
    \hfill 
    \begin{minipage}[t]{0.32\linewidth}
\caption{\raggedright Triangular charge-Kondo circuit. Metallic islands $A,B,C$ are coupled together and connected to external leads by QPCs. Gates control island charges $N_{\alpha}$ and transmissions $\tau_j$. 
}
        \label{fig:device}
    \end{minipage}
\end{figure}

In this \textit{Letter}, we consider a geometrically frustrated noncollinear configuration of three coupled islands, each coupled to its own lead -- see Fig.~\ref{fig:device}. In both weak and strong tunneling regimes, we show that this triangular structure supports a $\mathbb{Z}_3$ parafermionic critical point, connected to the cyclic permutation of three collective charge states. Depending on gate tuning, the parafermion can be shared among the three islands or controllably relocalized to any pair of islands. Scaling up, this suggests the basis for extended networks of coupled islands, where non-Abelian anyons can be realized and manipulated in a reconfigurable architecture, with potential quantum information applications~\cite{kam,mzm2,mzm3}.

\clearpage

\textbf{\textit{Device setup.--}}
The triangular charge-Kondo circuit illustrated
in Fig.~\ref{fig:device} consists of three
micron-scale metallic islands $\alpha=A,B,C$, with equal charging energies $E_C$ and effectively continuous single-particle spectra. In the spin-polarized quantum Hall regime, each island is coupled to an external lead through a tunable single-channel QPC, labeled $j=1,2,3$. Three other QPCs,
$j=4,5,6$, connect the islands $AB$, $BC$, and $CA$. 
Applied gate voltages control the island charges $N_{\alpha}$ and bare QPC transmissions $\tau_j=1-|r_j|^2$, where $r_j$ are the reflection amplitudes.

Each QPC couples two spinless electronic continua, one on either side of the contact. Contacts on the same metallic island $\alpha$ are treated as independent electronic reservoirs but share its total charge $N_{\alpha}$. The microscopic Hamiltonian is
$H=H_0+H_C+H_{\mathrm{QPC}}$, with
\begin{eqnarray}
H_0 &= \sum_{j=1}^{6}\sum_{\gamma=1}^{2}\sum_k
\epsilon_k^\pdag\: \psi^\dagger_{j\gamma k}\psi^\pdag_{j\gamma k},
\label{eq:H0}\\
H_{\mathrm{QPC}} &= \sum_{j=1}^{6}
\left[t_j\psi^\dagger_{j1}(0)\psi_{j2}(0)
+\mathrm{H.c.}\right],
\label{eq:HQPC}\\
H_C &= E_C\sum_{\alpha}(\delta\hat q_\alpha)^2
+\sum_{\alpha<\beta}V_{\alpha\beta}
\delta\hat q_\alpha\delta\hat q_\beta.
\label{eq:HC}
\end{eqnarray}
Here $\psi^\dagger_{j\gamma k}$ creates an electron of energy
$\epsilon_k$ in continuum $\gamma$ of QPC $j$, and
$\psi_{j\gamma}(0)$ is its local contact field. For $j=1,2,3$,
$\gamma=1$ ($2$) denotes lead (island) electrons; for $j=4,5,6$, $\gamma=1$ ($2$) denotes electrons on island $A$ ($B$), $B$ ($C$), and $C$ ($A$). The amplitude $t_j$ controls tunneling across contact $j$. It can be tuned from the weak-transmission `Kondo' regime to the weak-backscattering `quasiballistic' regime, both of which we examine in this work. 
We define $\delta\hat q_\alpha = \hat N_\alpha-N_\alpha^0-N_{g\alpha}$,
where $\hat N_\alpha$ counts electrons in all continua belonging to island $\alpha$, $N_\alpha^0$ is a reference occupation, and $N_{g\alpha}$ is a dimensionless gate-controlled charge. Mutual capacitive interactions are set by $V_{\alpha\beta}$. The relative energies of different island charge configurations are thus determined by $N_{g\alpha}$ and $V_{\alpha\beta}$. 


\textbf{\textit{Kondo regime.--}}
For small-to-intermediate QPC transmissions and $T\ll E_C$, we retain the two lowest states $|0\rangle_\alpha\equiv|N_\alpha^0\rangle$
and $|1\rangle_\alpha\equiv|N_\alpha^0+1\rangle$ on each island $\alpha$.
Defining $S_\alpha^+=|1\rangle_\alpha\langle0|$ and
$S_\alpha^z=(|1\rangle_\alpha\langle1|-|0\rangle_\alpha\langle0|)/2$
gives $\hat N_\alpha=N_\alpha^0+\tfrac12+S_\alpha^z$ in the projected
space. The charging Hamiltonian becomes, up to a constant,
\begin{eqnarray}
H_C^{\mathrm K}&=\sum_\alpha h_\alpha^\pdag S_\alpha^z
+\sum_{\alpha<\beta}V_{\alpha\beta}^\pdag S_\alpha^zS_\beta^z,
\label{eq:kondo_charging}
\end{eqnarray}
where $h_\alpha=2E_C\left(\tfrac12-N_{g\alpha}\right)
+\sum_{\beta\ne\alpha}V_{\alpha\beta}
\left(\tfrac12-N_{g\beta}\right)$. 
For simplicity we set $V_{\alpha\beta}\equiv V$. 
Following the charge-Kondo construction~\cite{Matveev_1991,ccd} 
we separate collective island charges from continuum
electronic excitations. 
Charge \textit{changes} due to QPC tunneling are represented by pseudospin ladder operators. Defining $b_j=\psi_{j1}^\dagger(0)\psi_{j2}(0)$, the projected
tunneling Hamiltonian becomes, 
\begin{eqnarray}
&H_{\rm QPC}^{\mathrm K}=t_1S_A^-b_1+t_2S_B^-b_2+t_3S_C^-b_3 \label{eq:kondo_tunneling}
\\
&+t_4S_A^+S_B^-b_4+t_5S_B^+S_C^-b_5 +t_6S_C^+S_A^-b_6+\mathrm{H.c.} \nonumber
\end{eqnarray}
Each tunneling event flips the island pseudospins and
scatters an electron between the continua.


\textbf{\textit{Isolated parafermion limits.--}}
We first consider two configurations in which a critical subsystem is exactly isolated from the rest of the circuit. Closing the external QPCs to the leads $t_1\!=\!t_2\!=\!t_3\!=\!0$, gives the isolated triangular triple dot. The sector with one excess electron, $Q=\sum_\alpha(\tfrac12+S_\alpha^z)=1$, contains the triplet $\vec{\phi}_{TCK}=(|100\rangle,|010\rangle,|001\rangle)$. These states are exchanged by tunneling at QPCs $4,5,6$. 
For $V_{\alpha\beta}\equiv V>0$ the bare electrostatic condition for stability of this configuration is $h_\alpha\equiv h$ and $0<h<V$. We call this the `triangular charge-Kondo' (TCK) model. The $Q\!=\!2$ sector is related by a particle-hole transformation and hosts the same physics, while $Q\!=\!0,3$ are trivial. 

Alternatively, we can obtain the double charge-Kondo (DCK) circuit~\cite{ccd,KBM1} by freezing out island $C$ and disconnecting it from the rest of the system by setting $t_3\!=\!t_5\!=\!t_6\!=\!0$. Its triple point at $h_A=h_B=V$ and $h_C>V$ involves the configurations 
$\vec{\phi}_{DCK}=(|000\rangle,|100\rangle,|010\rangle)$, which are exchanged  by tunneling at QPCs $1,4,2$.

For $T,T_K\ll\Delta_{\mathrm{ch}}$, with 
$\Delta_{\mathrm{ch}}$
the gap to other charge configurations and $T_K$ the Kondo scale, projection onto either triplet gives the same cyclic exchange model, 
\begin{eqnarray}
H_3&=H_0+\sum_{a=1}^3
\left[J_a X_{a,a+1}\mathcal{B}_a+{\rm H.c.}\right],
\label{eq:triplet_model}
\end{eqnarray}
where $X_{ab}=|a\rangle\langle b|$ interconverts the retained island charge configurations (indices are cyclic).
The electronic scattering terms are $(\mathcal B_1,\mathcal B_2,\mathcal B_3)
=(b_4,b_5,b_6)$ for TCK and $(b_1,b_4,b_2^\dagger)$ for DCK. Thus the same three-state model describes the rearrangement of charge among $A,B,C$, or charge transfer through $A,B$ and their leads. The critical point corresponds to degenerate triplet energies and equal effective couplings $J_a\equiv J$~\cite{Jnote,ccd,KBM1}.

At the critical point, we have an emergent $\mathbb{Z}_3$ parafermion, which we can construct explicitly in terms of our microscopic degrees of freedom by applying a generalized Emery-Kivelson (EK) mapping~\cite{EK1,FG1,KBM1}. 

First we bosonize with canonically normalized
chiral fields,
$\mathcal B_a\propto e^{i\sqrt{2}\phi_a}$, absorbing short-distance prefactors into the couplings~\cite{vonDelft1998}. With
$\phi_s=(\phi_1+\phi_2+\phi_3)/\sqrt{3}$ and
$\chi_a=\sqrt{2}(\phi_a-\phi_{a-1})/3$, we 
introduce the longitudinal deformation
$H_z=v_F\sum_aX_{aa}\partial_x\chi_a(0)$ where  $\phi_0\equiv\phi_3$ and $v_F$ is the Fermi velocity.
The EK transformation of $H'=H_3+H_z$ gives 
the effective Toulouse Hamiltonian in which only the field $\phi_s$ couples:
\begin{equation}
H_{\mathrm{EK}}\equiv U H' U^{\dagger}=H_0+
\left[J\Lambda e^{i\sqrt{2/3}\phi_s(0)}+\mathrm{H.c.}\right],
\label{eq:triplet_Toulouse}
\end{equation}
where we have used $U=\exp\left[-i\sum_aX_{aa}\chi_a(0)\right]$.

\begin{figure*}[t]
    \centering
    \begin{minipage}[t]{0.7\textwidth}
        \vspace{0pt}
        \includegraphics[width=\linewidth]{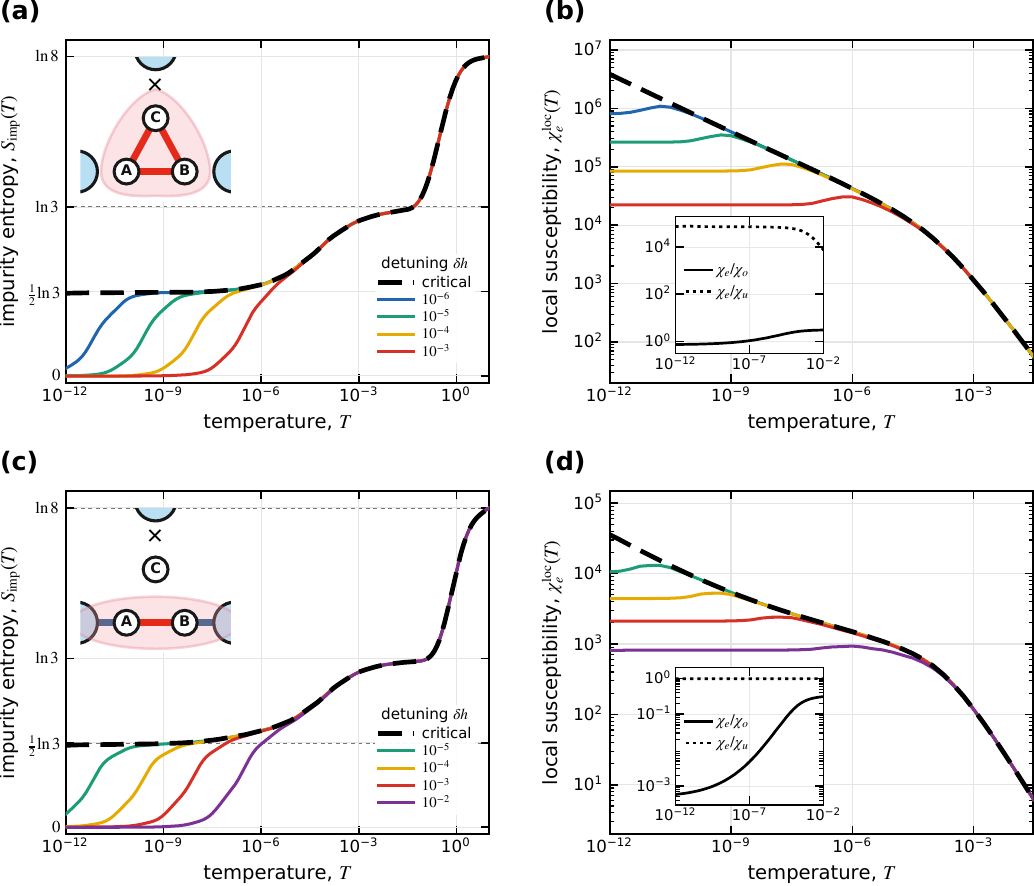}
    \end{minipage}
    \hfill 
    \begin{minipage}[t]{0.27\textwidth}
\caption{NRG results for the triangular charge-Kondo circuit model in the weak-tunneling regime. Here we close QPC 3 and eliminate the lead at island $C$. We initially set $t_j\!=\!0.2$ for all $j\!\ne\! 3$, $V\!=\!1$, and $D\equiv 1$ the electron half-bandwidth. (a,b) For configuration $ABC$ we set $\vec{h}\equiv (h_A,h_B,h_C)=(V/2,V/2,V/2)$ and then retune $t_4\simeq 0.19$ to realize the TCK critical point. 
(c,d) For configuration $AB$  we set $\vec{h}=(V,V,2V)$ and retune $t_4\simeq 0.17$ to realize the DCK critical point. In both cases, the critical point is shown as the black dashed lines, and $T_K\sim 10^{-4}$. 
Solid lines show the subsequent Fermi liquid crossover due to the gate detuning $\delta h$ applied to islands $A$ and $B$ as given in the legend, where $h_{A,B}\to h_{A,B}+\delta h$.
Panels (a,c) show the impurity entropy $S_{\rm imp}(T)$ while the main panels of (b,d) show the local charge susceptibility $\chi_e^{\rm loc}(T)$. Insets show the ratios $\chi_e^{\rm loc}/\chi_o^{\rm loc}$ and $\chi_e^{\rm loc}/\chi_u^{\rm loc}$.
}
        \label{fig:NRG}
    \end{minipage}
\end{figure*}

Here $\Lambda=X_{12}+X_{23}+X_{31}$ satisfies $\Lambda^3=1$ and $\Lambda^2=\Lambda^\dagger$. With $Z=X_{11}+\omega X_{22}+\omega^2X_{33}$ and $\omega=e^{2\pi i/3}$, we have the clock relation $\Lambda Z=\omega Z\Lambda$. The operators $\alpha_1=\Lambda$ and $\alpha_2=\Lambda Z$ therefore obey a $\mathbb{Z}_3$ parafermionic algebra $\alpha_i^3=1$ and $\alpha_1\alpha_2=\omega\alpha_2\alpha_1$.

Since $[H_{\rm EK},\alpha_1]=0$, one $\mathbb{Z}_3$ parafermion remains decoupled while its partner is screened. Diagonalizing $\Lambda$ gives three uncoupled $\nu=1/3$ boundary sine-Gordon models, corresponding to the potentials $2J\cos[\sqrt{2/3}\phi_s+2\pi m/3]$ with $m=0,\pm1$, consistent with a $\mathbb{Z}_3$ parafermionic structure~\cite{Fendleyz, Fendley1998}. The exact entropy change at the boundary~\cite{Fendleyz,AffleckLudwig1991BoundaryEntropy} gives
$S_{\mathrm{imp}}=\ln3\to\tfrac12\ln3$, which is as such a hallmark of the common low-temperature critical sector for both configurations. 

We note that the Toulouse deformation $H_z$ is RG-irrelevant at the critical fixed point~\footnote{The critical charge operators have scaling dimension $1/3$, so longitudinal density couplings have dimension $4/3$ and are RG-irrelevant}. By contrast, charge-state splittings and coupling anisotropies are relevant and remove the free parafermion on the lowest energy scales. For details, see \textit{End Matter} and~\cite{Supplement}.

In the idealized limits discussed above, the $\mathbb{Z}_3$ parafermion can be delocalized over all three islands, or localized on any pair, depending on gate tuning.  In the following, we show that this critical physics survives in the full circuit setup, and the quasiballistic limit.


\textbf{\textit{NRG results.--}}
We first examine the Kondo limit (Eqs.~\ref{eq:H0},\ref{eq:kondo_charging},\ref{eq:kondo_tunneling}) with the numerical renormalization group (NRG)~\cite{qmp3,qmp1,mitchell2014generalized}, see \textit{End Matter} for technical details. Our goal now is to directly validate the EK prediction and demonstrate the different physical embeddings of the same $\mathbb Z_3$ critical point within a single device. The QPCs connecting the three islands are kept open and all 8 charge states are retained. We also keep open QPCs $1,2$ which attach leads to islands $A$ and $B$. However, we close QPC $3$ to make the problem numerically tractable with NRG. Nevertheless, the reduced configuration allows us to test whether (i) the $ABC$ critical state is robust to coupling to leads via $1,2$; and (ii) the $AB$ critical state is robust to coupling to island $C$. 

A summary of our NRG results is presented in Fig.~\ref{fig:NRG}. Panels (a,b) are for the triangular $ABC$ case where the triplet of island charge states $\vec{\phi}_{TCK}$ is lowest. 
Panels (c,d) are for the $AB$ case with $\vec{\phi}_{DCK}$ instead the lowest triplet. In both cases, gate voltages and couplings are tuned to reach the critical point, shown as the black dashed lines. Solid lines correspond to gate detuning $\delta h$ away from the critical point.  
Left panels show the `impurity' (charge pseudospin) contribution to the entropy $S_{\rm imp}(T)$.  At high temperatures $T\gg \Delta_{ch}\sim \mathcal{O}(V)$ all $2^3=8$ charge states are accessible, giving a common $\ln(8)$ entropy in both cases. For $T\lesssim \Delta_{ch}$ the entropy is reduced to $\ln(3)$ corresponding to the lowest charge triplets $\vec{\phi}_{TCK}$ or $\vec{\phi}_{DCK}$. 
At the critical point, the entropy flows to $\ln(3)/2$ below the Kondo temperature $T_K$, corresponding to the unscreened residual $\mathbb{Z}_3$ parafermion~\cite{KBM1}. Small gate detuning $\delta h$ generates a Fermi liquid scale 
$T^*/T_K\sim |\delta h/T_K|^{3/2}$ 
below which the entropy is quenched and the parafermion is lost. The universal low-energy curves scaled in terms of $T/T_K$ or $T/T^*$ are identical; only scale prefactors differ between the $ABC$ and $AB$ cases. 
Panels (a,c) confirm that both criticalities can be accommodated in the same device, with $ABC$ persisting with finite lead couplings, and $AB$ surviving the coupling to island $C$.

The right panels (b,d) of Fig.~\ref{fig:NRG} show that the physical embedding of the critical overscreened triplet is different in the two settings, being distinguished by their  charge response. We consider
static local susceptibilities $\chi_\nu^{\mathrm{loc}}(T)$ of
$Q_u=S_A^z+S_B^z+S_C^z$, $Q_o=S_A^z-S_B^z$, and
$Q_e=S_A^z+S_B^z-2S_C^z$. The main panels of (b,d) plot $\chi_e^{\mathrm{loc}}(T)$, and both show a singular low-temperature response at the critical point $\sim (T/T_K)^{-1/3}$, consistent with the dimension-$1/3$ critical field of Eq.~\ref{eq:triplet_model}. 
However, the insets reveal contrasting response
modes. For $ABC$, the relative-charge susceptibilities greatly exceed the uniform response, consistent with predominantly charge-conserving rearrangements among the three islands. For $AB$, $\chi_e^{\mathrm{loc}} =\chi_u^{\mathrm{loc}}$ as expected when $C$ charge fluctuations are suppressed. Indeed, the low-$T$ renormalization giving $\chi_o^{\mathrm{loc}}\gg \chi_e^{\mathrm{loc}}$ indicates critically-enhanced anticorrelation of charge on islands $A$ and $B$.


\textbf{\textit{Quasiballistic regime.--}} 
We now restore QPC~3 and consider all six contacts near perfect transmission, $|r_j|\ll1$. This quasiballistic regime is directly relevant to experimental realizations since strong couplings are needed to boost the critical Kondo scales into the accessible temperature window~\cite{Pierre_2018,ccd}.  

Within the bosonization framework~\cite{vonDelft1998,Matveev1995,Furusaki1995b}, we define boundary fields $\varphi_j=[\varphi_{j,R}(0)-\varphi_{j,L}(0)]/2$ at each QPC $j$,
where $\varphi_{j,R/L}$ are chiral bosonic fields for the 
right- and left-moving electrons. 
Expanding Eq.~\ref{eq:HQPC} around perfect transmission gives a backscattering term, 
\begin{eqnarray}
H_b&=-\frac{D}{\pi}\sum_{j=1}^6 |r_j|\cos(2\varphi_j),
\label{eq:Hb}
\end{eqnarray}
where $D$ is an ultraviolet cutoff, and ballistic contributions are absorbed into the free Hamiltonian $H_0$, see \textit{End Matter}. The interaction Hamiltonian Eq.~\ref{eq:HC} simplifies in terms of the $\varphi_j$ fields, as shown in the SM~\cite{Supplement}, 
with three field combinations acquiring a charging mass while three remain gapless. Rotating to this basis with $\varphi=\bar{S}\Phi$ and integrating out the gapped modes leaves a description of the backscattering dynamics at energies below the charging scale in terms of the three remaining orthogonal fields, denoted $\boldsymbol\Phi=(\boldsymbol\Phi_\perp,\Phi_c)$~\cite{Supplement}. This leaves a 3d quantum Brownian-motion problem~\cite{Yi,qba,Yi1},
\begin{eqnarray}
H_b'=-\sqrt{DE_C}\:{\textstyle\sum_{j=1}^{6}}|\tilde{r}_j|
 \cos(\eta_j+\mathbf k_j\cdot\boldsymbol\Phi) \;,\label{eq:qb_projected_model} \\
\mathbf k_\alpha=(\mathbf v_\alpha,0),\qquad
\mathbf k_{\alpha\beta}
=\tfrac{1}{3}\left(\mathbf v_\beta-\mathbf v_\alpha,\:
2\sqrt3\right),
\label{eq:qb_wavevectors}
\end{eqnarray}
where $|\tilde{r}_j|\propto |r_j|$ and $\alpha=A,B,C$ labels QPCs $j=1,2,3$ while $\alpha\beta=AB,BC,CA$ labels QPCs $j=4,5,6$. The vectors $\mathbf v_\alpha$ are planar and satisfy $|\mathbf v_\alpha|^2=2$ and $\mathbf v_\alpha\cdot\mathbf v_\beta=-1$ for $\alpha\ne\beta$. Thus 
the outer wavevectors $\mathbf k_\alpha$ form an equilateral triangle and inner wavevectors $\mathbf k_{\alpha\beta}$ share a common $\Phi_c$ component. 
The gate phases are $\eta_\alpha=\pi(N_\Sigma+N_{g\alpha})/2$
and $\eta_{\alpha\beta}=\pi(N_{g\alpha}-N_{g\beta})/2$, with
$N_\Sigma=\sum_\alpha N_{g\alpha}$.
The retained fields have Ohmic Gaussian dynamics, with vertices of dimension $|\mathbf k_j|^2/4=1/2$. 

In the limits below, 
we consider the renormalized pinning effect when certain reflections dominate, and the ultimate fate of the remaining dynamical fields.


\textbf{\textit{ABC criticality.--}}
Here we consider embedding the TCK critical point within the full circuit. We take $|r_{1,2,3}|=r_o$ and $|r_{4,5,6}|= r_i$, with $r_o\gg r_i$ such that the outer QPCs flow under RG to the strong barrier regime first. 
The dominant outer potentials in Eq.~\ref{eq:qb_projected_model} are characterized by the angles  
$\theta_\alpha=\eta_\alpha+
\mathbf v_\alpha\cdot\boldsymbol\Phi_\perp$ for $\alpha=A,B,C$. It pins the two components of
$\boldsymbol\Phi_\perp$ while leaving the internal-cycle field
$\Phi_c$ free. Since $\sum_\alpha\mathbf v_\alpha=0$, the angles satisfy $\sum_\alpha\theta_\alpha=2\pi N_\Sigma$.
Away from half-integer $N_\Sigma$, the minimizing branch is 
$\theta_\alpha^*=\frac{2\pi}{3}(N_\Sigma-m)+2\pi\ell_\alpha$ 
where $m=\sum_\alpha\ell_\alpha\in\mathbb Z$ and $|N_\Sigma-m|<\tfrac12$. 
To leading order in $r_i$, integrating out fluctuations of the
pinned outer fields $\boldsymbol\Phi_\perp$ about this minimum gives the residual backscattering potential from the inner contacts~\cite{Supplement},
\begin{eqnarray}
\delta H_{ABC}=-\left[\mathcal R_{ABC}\:
 e^{2i\Phi_{c}/\sqrt{3}}+\mathrm{H.c.}\right],
\label{eq:qb_ABC_reduction}
\end{eqnarray}
where the effective coupling is,
\begin{eqnarray}
\mathcal R_{\rm ABC}&=\lambda_{\rm ABC} \sum_{\alpha\beta}
 e^{\frac{2\pi i}{3} (N_{g\alpha}-N_{g\beta}-\ell_\alpha+\ell_\beta)}.\label{effectivec1}
\end{eqnarray}
Here $\lambda_{\rm ABC}\propto\sqrt{DE_C}\,|\tilde r_i|
\left(\frac{\mu_o}{D}\right)^{1/6} >0$ are the renormalized inner-reflection coefficients, where $\mu_o\sim E_C|\tilde r_o|^2$ is the outer pinning mass. For further details see \textit{End Matter} and SM~\cite{Supplement}. 
The three contributions to the residual backscattering amplitude cancel when $N_{g\alpha}=1/3$ on all three islands (as well as for its particle-hole partner $N_{g\alpha}=2/3$), giving $\mathcal R_{\rm ABC}=0$. Importantly, the effective cyclic symmetry of the low-energy model protects the points where Eq.~\ref{eq:qb_ABC_reduction} vanishes.  
At these zeros, $\Phi_c$ remains unpinned. Along this mode, the smallest charge-conserving cycle transfers one electron across each inner QPC and shifts $\Phi_c$ by $\pi\sqrt3$. Together with its Gaussian dynamics, this fixes the elementary boundary vertex to have dimension $1/3$. 
The surviving mode and its allowed vertices coincide with the $\nu=1/3$ DCK critical theory~\cite{KBM,KBM1}, which is associated with a nontrivial residual entropy  of $\tfrac12\ln3$~\cite{fendley1994}. 
A nonzero $\mathcal R_{\rm ABC}$ grows under RG and ultimately pins $\Phi_{c}$ as well, driving the system towards a zero-entropy Fermi liquid state. 

Residual tunneling through the outer barriers is captured by instanton physics between pinned minima. The shortest instanton processes dominate, but here even these are found to be irrelevant at low energies~\cite{Supplement}. 
Pinning by the outer potentials is therefore stable, and the ABC critical state survives finite coupling to leads.


\textbf{\textit{AB criticality.--}}
We now examine embedding the DCK critical point in the full model
by taking $|r_5|=|r_6|=r_s\gg|r_j|$ for $j=1,2,3,4$.
The two dominant links to island $C$ independently pin their
corresponding angles in the effective potential
Eq.~\ref{eq:qb_projected_model} to integer multiples of $2\pi$.
Up to a constant, the remaining normalized field is
$\Phi_{\rm AB}=(\varphi_1+\varphi_2+\varphi_4)/\sqrt3$. 
QPC~3 has no projection onto this mode. Integrating out the pinned fluctuations again gives
Eq.~\eqref{eq:qb_ABC_reduction} to leading order, but now with
$\Phi_c\to\Phi_{\rm AB}$ and
$\mathcal R_{\rm ABC}\to\mathcal R_{\rm AB}$, where
\begin{equation}
\mathcal R_{\rm AB}=\lambda_4^{\rm AB}
 +\lambda_1^{\rm AB}e^{-2\pi iN_{gA}}
 +\lambda_2^{\rm AB}e^{2\pi iN_{gB}}.
\label{eq:qb_AB_amplitude}
\end{equation}
Here $\lambda_j^{\rm AB}\propto\sqrt{DE_C}\,|\tilde r_j|
(\mu_s/D)^{1/6}>0$ are the residual backscattering
coefficients from contacts $j=1,2,4$, where
$\mu_s\sim E_C|\tilde r_s|^2$ is the side-link pinning scale.

In the case of mirror symmetry where $N_{gA}=N_{gB}=n$ and
$\lambda_1^{\rm AB}=\lambda_2^{\rm AB}=\lambda_o$,
destructive interference gives $\lambda_4^{\rm AB}+2\lambda_o\cos(2\pi n)=0$. 
There are two solutions per gate period for
$0<\lambda_4^{\rm AB}<2\lambda_o$. For equal renormalized amplitudes, this then gives $\mathcal R_{\rm AB}=0$ at $n=1/3$ and $2/3$. At this point, $\Phi_{\rm AB}$ remains unpinned

The elementary processes through QPCs~1,4,2
move an electron along the lead-$A$-$B$-lead path, and shifts $\Phi_{\rm AB}$ by $\pi\sqrt3$ as in the DCK model~\cite{KBM,KBM1}. As with the $ABC$ case, the shortest instantons through the side barriers are likewise irrelevant here~\cite{Supplement}. 
Thus the tuned $AB$ critical state is stable to finite coupling to island $C$, including an open QPC~3.\\


\begin{figure}[t]
    \centering
    \includegraphics[width=0.9\linewidth]{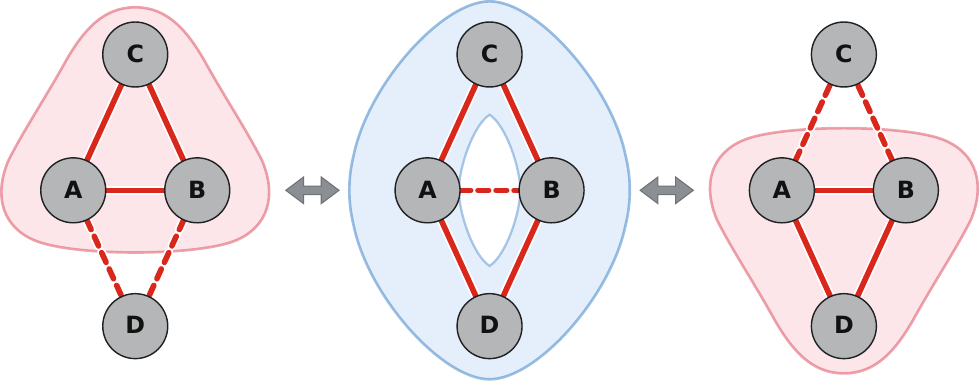}
    \caption{Networks of charge-Kondo islands, based on the triangular motif, support tunable Kondo anyons. A circuit containing four islands can host a $\mathbb{Z}_3$ parafermion localized on frustrated upper or lower triangles depending on gate tuning, while a $\mathbb{Z}_4$ parafermion can emerge in the 4-site ring.\\}
    \label{fig:4-dot}
\end{figure}

\noindent\textbf{\emph{Conclusion \& outlook for parafermion networks.}} 
We examined a triangular triple-dot circuit based on the hybrid metal-semiconductor charge-Kondo simulator architecture of Refs.~\cite{Pierre_2015,Pierre_2018,han2022fractional,piquard2023observing,ccd,entt,Sriram2026}. 
At weak coupling, two distinct idealized limits were identified in which a $\mathbb{Z}_3$ parafermion can be realized due to frustrated interactions. Both involve three degenerate collective charge configurations that are cyclically permuted by electronic tunneling at QPCs, but their physical embedding in the triangular structure differs. By tuning gate voltages, a parafermion on islands $ABC$ can be relocated to any pair. Importantly, we show that these critical parafermions survive in the full circuit setup of Fig.~\ref{fig:device}, including in the experimentally-relevant regime of strong QPC transmission. 

Our triangular structure may serve as the building block for a scalable network in which geometric frustration competes with Kondo screening to produce Kondo anyons with gate-tunable localization. Fig.~\ref{fig:4-dot} shows a simple extension to four islands, where a $\mathbb{Z}_3$ parafermion can be localized on either upper or lower triangles. Closing the $AB$ QPC forms a ring structure in which a $\mathbb{Z}_4$ parafermion can also be realized~\cite{KBM}. Extending further, a synthetic 2d lattice built up using this triangular motif may provide a reconfigurable platform in which multiple anyons could be separately realized and manipulated, brought into contact, and controllably coupled. Recently, it was shown in Ref.~\cite{karki2026coupled} that two critical Majoranas, each with quantum dimension $d=\sqrt2$, could be individually prepared in a two-island circuit, and subsequently coupled together without losing the combined $d^2$ degeneracy of the zero-mode NFL manifold. In a triangular network, $N$ copies of the $\mathbb{Z}_3$ parafermion studied here, each with $d=\sqrt3$, could be localized and coupled. The resulting $\ln(d^N)$ residual entropy could potentially be measured experimentally via the techniques demonstrated in Ref.~\cite{entt}. Ultimately, it may be possible to encode and manipulate quantum information using gate-tunable Kondo anyons~\cite{kam,boo} realized in such circuit networks.\\

\begin{acknowledgments}
\textit{\textbf{Acknowledgments.--}}
We thank David Goldhaber-Gordon and Frederic Pierre for enlightening discussions about practical implementation. 
A.K.M.~acknowledges funding support from Taighde \'Eireann - Research Ireland under Grant No.~24/FFP-P/12816 and through Rinn Quantum under grant No.~23/RC/12197. 

\vfill
\clearpage


\appendix
\section*{End Matter}
\label{sec:endmatter}

\setcounter{figure}{0}
\renewcommand{\thefigure}{E\arabic{figure}}
\setcounter{equation}{0}
\renewcommand{\theequation}{E.\arabic{equation}}
\setcounter{table}{0}
\renewcommand{\thetable}{E.\arabic{table}}


\textbf{\textit{Details of EK transformation.--}}
We use chiral fields satisfying $[\phi_a(x),\phi_b(y)]
=i\pi\delta_{ab}\operatorname{sgn}(x-y)$ and,
\begin{equation}
H_0=\frac{v_F}{4\pi}\sum_{a=1}^3
\int dx\,(\partial_x\phi_a)^2+H_{\mathrm{free}},
\end{equation}
where $\hbar=1$ and $H_{\mathrm{free}}$ describes the uncoupled bath modes which play no role here.

For independent channels, the fermion transfer operators
$\mathcal B_a$ contain disjoint pairs of Klein factors, which therefore commute with each another and with the overall bosonized Hamiltonian. In a fixed Klein sector their phases can be absorbed into constant shifts of $\phi_a$, giving
$\mathcal B_a\propto e^{i\sqrt{2}\phi_a}$ as used in the
main text. 

The fields $\chi_a$ defined in the main text satisfy,
\begin{equation}
\sum_a\chi_a=0,
\qquad
\chi_a-\chi_{a+1}
=\sqrt{2}\phi_a-\sqrt{\frac{2}{3}}\phi_s.
\end{equation}
The EK rotation then acts on exchange vertices as,
\begin{equation}
U\left[X_{a,a+1}e^{i\sqrt{2}\phi_a(0)}\right]U^\dagger
=X_{a,a+1}e^{i\sqrt{2/3}\phi_s(0)} \;,
\end{equation}
while the kinetic term gives,
\begin{equation}
UH_0U^\dagger
=H_0-v_F\sum_aX_{aa}\partial_x\chi_a(0)
+\mathrm{const}.
\end{equation}
The resulting longitudinal term is canceled by $H_z$. 
With a common short-distance cutoff, the remaining shifts are proportional to the identity.
The coefficient vectors of all three $\chi_a$ have the
same squared norm, $4/9$. This gives Eq.~\ref{eq:triplet_Toulouse} up to a constant. The two field combinations orthogonal
to $\phi_s$ remain free.

To assess stability away from the Toulouse point, 
we define $Q_a=X_{aa}-1/3$, which are as such linear
combinations of $Z$ and $Z^\dagger$. These interconvert the
three sine-Gordon sectors, and change the pinned boundary field by
$2\pi/\sqrt{6}$ at the low-energy fixed point. Thus $Q_a$ has scaling dimension $1/3$, as with the DCK model~\cite{KBM1}. 
Residual longitudinal couplings, which involve
$Q_a\partial_x\chi_b$, have dimension $1/3+1=4/3$,
and are irrelevant. This establishes local stability to the deformation $H_z$. Any splitting of the triplet states or coupling anisotropies are however relevant and must be tuned away.

The free parafermion identified above is in the EK basis. In the original basis, the operator $U^\dagger\Lambda U$ is dressed, rather than being a bare permutation of charges. 


\textbf{\emph{NRG implementation \& observables.--}}
We solve the Kondo-limit Hamiltonian $H_K=H_0+H_C^K+H_{\rm QPC}^K$ given by Eqs.~\ref{eq:H0},\ref{eq:kondo_charging},\ref{eq:kondo_tunneling} using NRG. With QPC 3 closed and the corresponding lead attached to island $C$ removed, we have an effective 5-channel quantum impurity model (ten spinless continua either side of the 5 remaining QPCs, and three local charge pseudospin-$\tfrac{1}{2}$ impurity degrees of freedom for the three islands). Such a model is still too complex to treat directly, and so we exploit full non-Abelian symmetries~\cite{qmp2,weichselbaum2012non} and also employ the `interleaved' iNRG method~\cite{mitchell2014generalized,stadler2016interleaved}. 

Taking particle-hole symmetric flat bands of half-width $D\equiv 1$ for the electronic continua, we have $[SU(2)_c]^5$ for the charge isospin of the 5 channels. We also have three $U(1)$ symmetries related to spin. With:
\begin{eqnarray}
q_j&=\frac{1}{2}\sum_k
\left(
\psi_{j1k}^{\dagger}\psi_{j1k}^\pdag
-\psi_{j2k}^{\dagger}\psi_{j2k}^\pdag
\right),
\qquad j\ne 3 \qquad 
\label{eq:bath_charges}
\end{eqnarray}
the conserved charges are defined as $\mathcal Q_A = S_A^z+q_1-q_4+q_6$, $\mathcal Q_B = S_B^z+q_2+q_4-q_5$ and $\mathcal Q_C = S_C^z+q_5-q_6$, so that $[H,\mathcal Q_\alpha]=0$. 
The NRG Hamiltonians are block diagonal in the five charge-isospin quantum numbers and the three conserved charges $\mathcal Q_\alpha$. Using the Wigner-Eckart theorem, we work with reduced matrix elements to avoid keeping track of different isospin projections for states in the same multiplet~\cite{qmp2}. Each electronic channel contributes four local states at each NRG iteration, but these are organized into one charge-isospin doublet and two singlets within this symmetry setting. The $2^{10}=1024$ states of a complete Wilson shell therefore comprise only $3^5=243$ irreducible multiplets.

We further reduce the complexity by interleaving the Wilson chains of different channels into a single generalized Wilson chain~\cite{mitchell2014generalized} within the iNRG method~\cite{stadler2016interleaved}, and performing intermediate Fock-space truncations between substeps. Specifically, we add channels $j=1,2$ together in substep 1, then channels $j=5,6$ in substep 2, and finally channel $j=4$ in the final substep to complete that Wilson shell before going on to the next NRG iteration. This construction allows us to preserve exact $AB$ mirror-symmetry, which we make use of to realize the critical physics by setting $t_1=t_2$, $t_5=t_6$, $h_A=h_B$. 

We use an NRG discretization parameter $\Lambda=6$ and an energy-based truncation which typically retains around $N_k \sim  10^6$ states per iNRG substep. We verified that the reported physical quantities were converged with respect to $N_k$ and $\Lambda$. 

The thermodynamic entropy was computed directly using the complete Anders-Schiller basis across all iNRG substeps~\cite{anders2006spin}. We report the impurity contribution $S_{\rm imp}(T)=S_{\rm full}(T)-S_{\rm bath}(T)$ where $S_{\rm bath}$ was obtained from an independent iNRG run with the same parameters but without the island pseudospins. The results of two different $z$-shift calculations were then averaged~\cite{vzitko2009energy}. For display purposes we smoothed the data to remove oscillating discretization artifacts.

We also calculate the static local responses of the uniform,
odd and even island charge combinations, 
$Q_u = S_A^z+S_B^z+S_C^z$, $Q_o = S_A^z-S_B^z$, $Q_e = S_A^z+S_B^z-2S_C^z$.  
For a probe perturbation $H_B=-\sum_\nu B_\nu Q_\nu$,
the corresponding susceptibilities are
\begin{eqnarray*}
\chi_\nu^{\mathrm{loc}}(T)
\equiv \left.\frac{\partial\langle Q_\nu\rangle_T}
{\partial B_\nu}\right|_{\mathbf B=0} = \int_0^\beta d\tau\,
\langle\delta Q_\nu(\tau)\delta Q_\nu(0)\rangle_T
\end{eqnarray*}
where $\nu=u,o,e$, $\beta=1/T$, $\delta Q_\nu=Q_\nu-\langle Q_\nu\rangle_T$, and $Q_\nu(\tau)=e^{\tau H_K}Q_\nu e^{-\tau H_K}$. Thermal averages are taken with the full five-contact
Kondo Hamiltonian. We evaluate their Lehmann representation using the full density matrix NRG~\cite{fdmnrg}. With the definitions above, the bare (unrenormalized) $ABC$ triplet $\vec{\phi}_{TCK}$ would give $\chi_u^{\rm loc}=0$ and $\chi_e^{\rm loc}=3\chi_o^{\rm loc}$. By contrast, the $AB$ triplet $\vec{\phi}_{DCK}$ gives $\chi_e^{\rm loc}=\chi_u^{\rm loc}$ and $\chi_o^{\rm loc}=3\chi_e^{\rm loc}$. In the full coupled system these values get renormalized at low $T$. 

In the main text, we defined the critical Kondo scale $T_K$ via the entropy as $S_{\rm imp}(T_K)=\tfrac{3}{4}\ln(3)$. Similarly, the Fermi liquid scale $T^*$ is defined via $S_{\rm imp}(T^*)=\tfrac{1}{4}\ln(3)$. In both $ABC$ and $AB$ configurations, we tuned $t_4$ to critical point by minimizing $T^*$. The gate detuning perturbation $\delta h$ was introduced via $h_A=h_B=V/2+\delta h$ (but with $h_C=V/2$ unchanged) in the $ABC$ configuration; and $h_A=h_B=V+\delta h$ (but with $h_C=2V$ unchanged) in the $AB$ configuration.


\textbf{\textit{Details of quasiballistic analysis.--}}
We employ an effective 1d description of the system in terms of chiral right/left moving fields $\varphi_j=(\varphi_{j, \rm R}-\varphi_{j, \rm L})/2$ with commutation relation 
$[\varphi_{j,R/L}(0,t),\varphi_{j,R/L}(0,t')]=-i\pi\operatorname{sgn}(t-t')$. 
We can then write Eq.~\ref{eq:H0} as~\cite{WE1, Kane_1992, Furusaki1995b},
\begin{equation}
H_0=\frac{v_{\rm F}}{4\pi}\sum_{j=1}^{6}\int dx
\left[
(\partial_x\varphi_{j,R})^2+
(\partial_x\varphi_{j,L})^2
\right].\label{EM1}
\end{equation}
For common $V$, the charging Hamiltonian Eq.~\ref{eq:HC}  is,
\begin{equation}
\begin{aligned}
H_C&=\delta\hat{\mathbf q}^{\,T}K\,\delta\hat{\mathbf q},\\
K&=\left(E_C-\frac V2\right)I_3
  +\frac V2\,\mathbf1\mathbf1^T,
\end{aligned}
\label{eq:em_charging_matrix}
\end{equation}
where $\delta\hat{\mathbf q}=(\delta\hat q_A,\delta\hat q_B,
\delta\hat q_C)^T$ and $\mathbf1=(1,1,1)^T$.
The uniform mode $\mathbf1/\sqrt3$ has eigenvalue $E_+=E_C+V$,
while the two modes orthogonal to $\mathbf1$ have
eigenvalue $E_-=E_C-V/2$. We assume $0\leq V<2E_C$, so that
all three charging stiffnesses are positive.
The same orthogonal transformation $S$ specified in the
SM~\cite{Supplement} diagonalizes $H_C$ to give,
\begin{equation}
\begin{aligned}
\pi^2H_C={}&4E_-\left[
\Phi_1-\frac{\pi(2N_{gA}-N_{gB}-N_{gC})}{2\sqrt6}
\right]^2\\
&+4E_-\left[
\Phi_2-\frac{\pi(N_{gB}-N_{gC})}{2\sqrt2}
\right]^2\\
&+E_+\left[\Phi_3-\frac{\pi N_\Sigma}{\sqrt3}\right]^2,
\end{aligned}
\label{eq:em_charging_modes}
\end{equation}
where $N_\Sigma=\sum_\alpha N_{g\alpha}$. Note that $\Phi_{4,5,6}$ do not appear in $H_C$. Also, the transformation matrix $S$ does not depend on $V$.  Therefore the gate phases $\eta_j$ and wavevectors $\mathbf k_j$ of the projected backscattering Hamiltonian are $V$-independent. At energies below the charging scale, $V$ enters only through the proportionality constant connecting the effective $|\tilde r_j|$ to the bare $|r_j|$. Our physical conclusions are therefore unaffected by setting $V=0$, which we do now for simplicity.


The amplitudes in Eqs.~\eqref{eq:qb_ABC_reduction}
and~\eqref{eq:qb_AB_amplitude} include fluctuations about
the selected pinned minima. For the two orthonormal pinned
eigenmodes $x_a$, the zero-temperature Gaussian action is given by,
\begin{equation}
\mathcal S_{\rm pin}=\frac{1}{\pi}\sum_{a=1}^{2}
\int_{-D}^{D}\frac{d\omega}{2\pi}
\bigl(|\omega|+\mu_a\bigr)|x_a(\omega)|^2,
\label{eq:em_qb_piN_{gA}ction}
\end{equation}
where $\mu_a$ are renormalized masses and the
$D$ is the high-energy cutoff. 
Thus $\langle x_a^2\rangle=\tfrac12\ln[(D+\mu_a)/\mu_a]$.
With $k_{ja}^{\rm pin}$ denoting the components of
$\mathbf k_j$ along these modes, we can average the pinned part of the vertex to obtain,
\begin{equation}
F_j=\left\langle e^{i\sum_a k_{ja}^{\rm pin}x_a}\right\rangle
=\prod_{a=1}^{2}
\left(\frac{\mu_a}{D+\mu_a}\right)^{(k_{ja}^{\rm pin})^2/4}.
\label{eq:em_qb_mass_factor}
\end{equation}
For every active contact we have 
$\sum_a(k_{ja}^{\rm pin})^2=2/3$. This gives a common
pinning power of $1/6$.

For ABC, both masses equal $\mu_o$, so the common factor is $F_{\rm ABC}\simeq(\mu_o/D)^{1/6}$.
For AB, the masses are $\mu_s$ and $3\mu_s$, giving
$F_{1,2}\simeq3^{1/24}(\mu_s/D)^{1/6}$ and $F_4\simeq3^{1/6}(\mu_s/D)^{1/6}$. 
The corresponding coefficients are
$\lambda_{\rm ABC}=\tfrac12\sqrt{DE_C}\,|\tilde r_i|F_{\rm ABC}$
and
$\lambda_j^{\rm AB}=\tfrac12\sqrt{DE_C}\,|\tilde r_j|F_j$
for $j=1,2,4$. These factors help identify the matching energy scales, which determine which couplings flow to the strong barrier regime first. 
However, we note that these numerical prefactors are Gaussian estimates, and may be renormalized by nonlinear fluctuations. In particular, equal bare reflections need not give equal $\lambda_j^{\rm AB}$.

An instanton displacement $\delta\boldsymbol\Phi$ between pinned minima has dimension $\Delta_{\rm inst}=|\delta\boldsymbol\Phi|^2/\pi^2$. 
In our case, the shortest displacements are
\begin{equation}
\begin{aligned}
\delta\boldsymbol\Phi&=\frac{2\pi}{3}
(\mathbf k_A-\mathbf k_B)
&&\text{(ABC, fixed $m$)},\\
\delta\boldsymbol\Phi&=\frac{2\pi}{3}
(2\mathbf k_{BC}-\mathbf k_{CA})
&&\text{(AB)}.
\end{aligned}
\label{eq:em_qb_shortest_instantons}
\end{equation}
Both have squared length $8\pi^2/3$ and so 
$\Delta_{\rm inst}=8/3>1$. Such processes are therefore irrelevant at the critical fixed point (and higher-order processes are even more strongly irrelevant). For ABC, one can neglect interbranch tunneling processes provided there is a sufficiently large interbranch splitting. Fortunately, branch degeneracies are avoided at $N_\Sigma=1,2$ where the critical point arises~\cite{Supplement}.

\textbf{\textit{Connection with other models.--}} The four-channel Kondo model of Ref.~\cite{varr} and the topological Kondo effect proposed in Ref.~\cite{beri2012,*galpin2014conductance} also found a residual $\tfrac{1}{2}\ln3$ impurity entropy, and identified a diverging susceptibility $\chi_{\rm imp}\sim T^{-1/3}$. 
Possible connections between these critical points are left for future exploration.

\clearpage

\end{acknowledgments}

\vfill

\clearpage
\widetext
\setcounter{equation}{0}
\setcounter{figure}{0}
\setcounter{table}{0}
\setcounter{section}{0}
\setcounter{secnumdepth}{3}
\makeatletter
\@removefromreset{equation}{section}
\renewcommand{\theequation}{S\arabic{equation}}
\renewcommand{\thefigure}{S\arabic{figure}}
\renewcommand{\bibnumfmt}[1]{[S#1]}
\renewcommand{\citenumfont}[1]{S#1}
\renewcommand{\thesection}{S\arabic{section}}
\makeatother
\begin{center}
{\Large\textbf{Supplemental Material}}
\end{center}
In this supplemental material, we outline the derivation of the key equations used in the main text.
\section{General framework in the quasi-ballistic regime}
The physical device under consideration is effectively two-dimensional, so a point may be represented by $r=(x,y)$. However, each quantum Hall edge is a one-dimensional curve. Consequently, the fields describing QPCs implemented by a pair of counter-propagating edge states require only a single longitudinal coordinate along the curve. For edge pair $j$, we introduce a longitudinal parameter $s_j$ and an embedding $R_j(s_j)=(x_j(s_j),y_j(s_j))$, with $R_j(0)=R_{\mathrm{QPC},j}$. Thus, each QPC can have its own coordinate value $s_j=0$, even though the six QPCs are located at different physical coordinates $(x,y)$. Indeed, for an island labeled $\alpha$ and an edge pair $j$ touching the $\alpha$-th island, we choose the coordinate $s_j$ such that the right (R) mode propagates in the $+s_j$ direction and the left (L) mode in the $-s_j$ direction, with the QPC at $s_j=0$. In the semi-infinite approximation, the portion of the pair associated with island $\alpha$ is either $(-\infty,0]$ or $[0,\infty)$. With these conventions, all calculations can be projected onto one-dimensional space, with each QPC located at the origin. Hence, in what follows, we suppress the implicit two-dimensional notation and write everything in a general one-dimensional formalism. The correctness of this formalism can be verified by examining the explicit conservation of charge at the internal links (QPCs). Therefore, with the QPCs labeled by $j=1, 2,\cdots, 6$ and the islands labeled by $\alpha= A, B, C$, the Hamiltonian terms describing the kinetic part of non-interacting spinless electrons in the six inter-spaced QPCs, the total charging energy of the device, and the weak backscattering in each QPC take the general forms given in Eqs.~\eqref{EM1}, ~\eqref{eq:em_charging_matrix}, and~\eqref{eq:Hb}, respectively.

The charging energy Hamiltonian~\eqref{eq:em_charging_matrix} can be written as
\begin{align}
H_C=\frac{E_C}{\pi^2}\Bigg[\varphi^{\mathsf{T}}M \varphi+2\pi N^{\mathsf{T}}\varphi+\pi^2\left(N_{gA}^2+N^2_{gB}+N^2_{gC}\right)\Bigg],
\end{align}
where we have set $V=0$ and introduced the matrices
\begin{align}
M=\left(
\begin{array}{cccccc}
 1 & 0 & 0 & -1 & 0 & -1 \\
 0 & 1 & 0 & -1 & 1 & 0 \\
 0 & 0 & 1 & 0 & -1 & -1 \\
 -1 & -1 & 0 & 2 & -1 & 1 \\
 0 & 1 & -1 & -1 & 2 & 1 \\
 -1 & 0 & -1 & 1 & 1 & 2 \\
\end{array}
\right),\quad \varphi=\left(
\begin{array}{c}
 \varphi _1 \\
 \varphi _2 \\
 \varphi _3 \\
 \varphi _4 \\
 \varphi _5 \\
 \varphi _6 \\
\end{array}
\right),\quad N=\left(
\begin{array}{c}
 N_{gA} \\
 -N_{gB} \\
 -N_{gC} \\
 N_{gB}-N_{gA} \\
 N_{gC}-N_{gB} \\
 N_{gC}-N_{gA} \\
\end{array}
\right).
\end{align}
We then apply the orthogonal transformation $\varphi=S\Phi$ to the original fields, with
\begin{align}
S=\left(
\begin{array}{cccccc}
 -\frac{1}{\sqrt{6}} & 0 & -\frac{1}{\sqrt{3}} & \frac{1}{\sqrt{3}} & -\frac{1}{2 \sqrt{6}} & \frac{1}{2 \sqrt{2}} \\
 -\frac{1}{2 \sqrt{6}} & \frac{1}{2 \sqrt{2}} & \frac{1}{\sqrt{3}} & 0 & -\frac{\sqrt{\frac{3}{2}}}{2} & \frac{1}{2 \sqrt{2}} \\
 -\frac{1}{2 \sqrt{6}} & -\frac{1}{2 \sqrt{2}} & \frac{1}{\sqrt{3}} & \frac{1}{\sqrt{3}} & \frac{1}{\sqrt{6}} & 0 \\
 \frac{\sqrt{\frac{3}{2}}}{2} & -\frac{1}{2 \sqrt{2}} & 0 & 0 & 0 & \frac{1}{\sqrt{2}} \\
 0 & \frac{1}{\sqrt{2}} & 0 & 0 & \frac{\sqrt{\frac{3}{2}}}{2} & \frac{1}{2 \sqrt{2}} \\
 \frac{\sqrt{\frac{3}{2}}}{2} & \frac{1}{2 \sqrt{2}} & 0 & \frac{1}{\sqrt{3}} & -\frac{1}{2 \sqrt{6}} & -\frac{1}{2 \sqrt{2}} \\
\end{array}
\right),
\end{align}
in order to express $H_C$ in a more appealing form
\begin{align}
H_C
&=\frac{E_C}{\pi^2}\Bigg(4\left[\Phi_1{-}\frac{\pi}{2\sqrt6}\left(2 N_{gA}{-}N_{gB}{-}N_{gC}\right)\right]^2+4\left[\Phi_2{-}\frac{\pi}{2\sqrt2}\left(N_{gB}{-}N_{gC}\right)\right]^2+\left[\Phi_3{-}\frac{\pi}{\sqrt3}\left(N_{gA}{+}N_{gB}{+}N_{gC}\right)\right]^2\Bigg).
\end{align}
From the above Hamiltonian, we observe that only three orthogonal combinations, $\Phi_1$, $\Phi_2$, and $\Phi_3$, are affected by the charging energy, while the remaining combinations do not enter the charging-energy dynamics. Consequently, when the charging energy is the largest energy scale in the problem, these three fields are effectively gapped by it and can therefore be integrated out. The remaining three gapless modes, $\Phi_4$, $\Phi_5$, and $\Phi_6$, then govern the low-energy dynamics of the system. After integrating out the gapped modes and transferring the gate-voltage dependence from the charging-energy Hamiltonian to the boundary backscattering, we finally arrive at an effective backscattering Hamiltonian of the form
\begin{align}
H'_{ b}=-\sqrt{D E_C}&\Bigg(|\tilde{r}_1|\cos \left[\frac{\pi}{2}  \left(2 N_{gA}+N_{gB}+N_{gC}\right)+\frac{\Phi _5}{\sqrt{6}}-\frac{\Phi _6}{\sqrt{2}}-\frac{2 \Phi _4}{\sqrt{3}}\right]\nonumber\\
&+|\tilde{r}_2|\cos \left[\frac{\pi}{2}   \left(N_{gA}+2 N_{gB}+N_{gC}\right)-\sqrt{\frac{3}{2}} \Phi _5+\frac{\Phi _6}{\sqrt{2}}\right]\nonumber\\
&+|\tilde{r}_3|\cos \left[\frac{\pi}{2}   \left(N_{gA}+N_{gB}+2 N_{gC}\right)+\frac{2 \Phi _4}{\sqrt{3}}+\sqrt{\frac{2}{3}} \Phi _5\right]\nonumber\\
&+|\tilde{r}_4|\cos \left[\frac{\pi}{2}   \left(N_{gA}-N_{gB}\right)+\sqrt{2} \Phi _6\right]\nonumber\\
&+|\tilde{r}_5|\cos \left[\frac{\pi}{2}   \left(N_{gB}-N_{gC}\right)+\sqrt{\frac{3}{2}} \Phi _5+\frac{\Phi _6}{\sqrt{2}}\right]\nonumber\\
&+|\tilde{r}_6|\cos \left[\frac{\pi}{2}   \left(N_{gA}-N_{gC}\right)+\frac{2 \Phi _4}{\sqrt{3}}-\frac{\Phi _6}{\sqrt{2}}-\frac{\Phi _5}{\sqrt{6}}\right]\Bigg),\label{sm1}
\end{align}
where we have introduced new parameters
\begin{align}
&|\tilde{r}_j|=\frac{\sqrt{4^{1/3}e^\gamma}}{\pi}|r_j|,\quad j=1, 2, 3\nonumber\\
& |\tilde{r}_j|=\frac{2^{2/3}\sqrt{4^{1/3}e^\gamma}}{\pi}|r_j|,\quad j=4, 5, 6,
\end{align}
with $\gamma$ denoting the Euler-Mascheroni constant. Performing an additional orthogonal transformation
\begin{align}
\left(
\begin{array}{c}
 \Phi _a \\
 \Phi _b \\
 \Phi _c \\
\end{array}
\right)=S'\left(
\begin{array}{c}
 \Phi _4 \\
 \Phi _5 \\
 \Phi _6 \\
\end{array}
\right),\quad S'=\left(
\begin{array}{ccc}
 \sqrt{\frac{2}{3}} & \frac{1}{\sqrt{3}} & 0 \\
 -\frac{\sqrt{2}}{3} & \frac{2}{3} & -\frac{1}{\sqrt{3}} \\
 -\frac{1}{3} & \frac{\sqrt{2}}{3} & \sqrt{\frac{2}{3}} \\
\end{array}
\right),
\end{align}
the above backscattering Hamiltonian~\eqref{sm1} can be brought into the form presented in equation~\eqref{eq:qb_projected_model} of the main text, with the vectors 
$\mathbf{v}_A=\left(-1/\sqrt2, \sqrt{3/2}\right)$, $\mathbf{v}_B=\left(-1/\sqrt2, -\sqrt{3/2}\right)$, $\mathbf{v}_C=\left(\sqrt2, 0\right)$, and $\mathbf{\Phi}_\perp=\left(\Phi_a, \Phi_b\right)$. Using $S$ and $S'$, the transformation $\bar S=S\:[I_3\otimes S'^T]$ presented in the main text can be constructed.

\section{Weakly tunneling outer channels: ABC criticality}
To investigate the regime in which the outer channels are operated into strongly backscattering regime, we consider the symmetric case $|\tilde{r}_j|=r_0$, $j=1, 2, 3$ and $|\tilde{r}_j|=r_i$, $j=4, 5, 6$, and re-express the backscattering Hamiltonian~\eqref{eq:qb_projected_model} as $H'_{b}=H'_{\rm bo}+H'_{\rm bi}$ with
\begin{align}
H'_{\rm bo}=-r_o\sqrt{D E_C}&\Bigg(\cos \left[\frac{\pi}{2}  \left(2 N_{gA}+N_{gB}+N_{gC}\right)-\frac{\Phi_a}{\sqrt2}+\sqrt{\frac{3}{2}}\Phi_b\right]\nonumber\\
&+\cos \left[\frac{\pi}{2}   \left(N_{gA}+2 N_{gB}+N_{gC}\right)-\frac{\Phi_a}{\sqrt2}-\sqrt{\frac{3}{2}}\Phi_b\right]\nonumber\\
&+\cos \left[\frac{\pi}{2}   \left(N_{gA}+N_{gB}+2 N_{gC}\right)+\sqrt2 \Phi_a\right]\Bigg),
\end{align}
and
\begin{align}
H'_{\rm bi}=-r_i\sqrt{D E_C}&\Bigg(\cos \left[\frac{\pi}{2}   \left(N_{gA}-N_{gB}\right)-\sqrt{\frac{2}{3}}\Phi_b+\frac{2}{\sqrt3}\Phi_c\right]\nonumber\\
&+\cos \left[\frac{\pi}{2}   \left(N_{gB}-N_{gC}\right)+\frac{\Phi _a}{\sqrt{2}}+\frac{\Phi _b}{\sqrt{6}}+\frac{2 }{\sqrt{3}}\Phi _c\right]\nonumber\\
&+\cos \left[\frac{\pi}{2}   \left(N_{gC}-N_{gA}\right)-\frac{\Phi _a}{\sqrt{2}}+\frac{\Phi _b}{\sqrt{6}}+\frac{2 }{\sqrt{3}}\Phi _c\right]\Bigg).
\end{align}
From above decomposition, we see that the vector of outer QPCs backscattering from an $A_2$, i.e., equal length vectors in a 2D plane, separated by 120 degrees, whose sum is zero, triangular set entirely in the $\left(\Phi_a, \Phi_b\right)$ plane, while that of the inner QPCs from another triangular set in $\left(\Phi_a, \Phi_b\right)$ with having a common $\Phi_c$ mode.

We note that since all terms in $H'_{\rm b}$ are of scaling dimension $1/2$, in the limit $r_o\gg r_i$, the outer QPCs first reach the strong potential limit, which pins the relevant bosonic fields $\Phi_a=\Phi_a^*$ and $\Phi_b=\Phi^*_b$ compatible with the gate voltages. To find this pinning energy scale, we follow a simple renormalization group (RG) procedure as follows. At the running cutoff $E$, we define the dimensionless boundary coupling $y(E)$. At the tree level of RG, we have $\frac{dy}{d\ell}=\left(1-\delta\right) y,\quad \ell=\ln\left(\frac{D}{E}\right)$, where $\delta$ is the scaling dimension of the backscattering Hamiltonian. The solution is $y(E)=y(D)\sqrt{\frac{D}{E}}$. For the outer QPCs, we have $y_o(E)=\frac{r_o\sqrt{D E_C}}{D}\sqrt{\frac{D}{E}}$. We then define the strong coupling scale $E^*_o$ such that $y_o(E^*_o)\sim 1$, giving $E^*_o\sim r_o^2 E_c$.

We denote the arguments of the three cosine terms in $H_{\rm bo}'$ by $\theta_j$ ($j=1,2,3$), which satisfy the constraint $\sum_j\theta_j=2\pi N_\Sigma$, where $N_\Sigma=N_{gA}+N_{gB}+N_{gC}$. The outer potential $H_{\rm bo}'$ reaches its minimum value, $V_{\rm o,min}=-3r_0\sqrt{E_cD}$, when $\theta_j=2\pi l_j$, with $l_j\in\mathbb{Z}$. Combining this global minimum condition with the constraint on the sum of the phases yields $N_\Sigma=L$, where $L=l_1+l_2+l_3$. For gate voltage ranges $0\leq N_j\leq 1$, the allowed integer values of $L$ are $L\in\{0,1,2,3\}$. By directly minimizing $H_{\rm bo}'$ with respect to $\Phi_a$ and $\Phi_b$, we obtain the stationary points as 
\begin{align}
 \Phi_a^*=\frac{\pi}{6\sqrt2}\left[(N_{gA}+N_{gB}-2N_{gC})+4(2l_3-l_1-l_2)\right],\quad   \Phi_b^*=\frac{\pi}{2\sqrt6}\left[(N_{gB}-N_{gA})+4(l_1-l_2)\right].
\end{align}
When fluctuations around the stationary points, $\Phi_a=\Phi_a^{*}+\delta\Phi_a$ and $\Phi_b=\Phi_b^{*}+\delta\Phi_b$, are included and the global minimization constraint $N_\Sigma=L$ is imposed, the strongly backscattering outer-sector Hamiltonian generates a term of the form ${\mathcal H}'_{\rm bo} = H_{\rm cont} + \frac{\mu_o}{\pi} \left(\delta\Phi_a^2+\delta\Phi_b^2\right)$, with $\mu_o \sim r_o^2 E_C$ and $H_{\rm cont}$ being a constant term. In terms of the effective action, fluctuations around the stationary minima then generate a quadratic action of the form
\begin{align}
S_0^{ab}=\frac{1}{\beta}\sum_n\frac{1}{\pi}\left(|\omega_n|+\mu_0\right)\left(\delta\Phi_a^2+\delta\Phi_b^2\right),\label{sm2}
\end{align}
where $\beta$ is the inverse temperature. Including fluctuations around the stationary points, the inner part of the backscattering Hamiltonian becomes
\begin{align}
H'_{\rm bi}=-r_i\sqrt{D E_C}&\Bigg(\cos \left[\frac{2\pi}{3}   \left(\tilde{N}_{gA}-\tilde{N}_{gB}\right)-\sqrt{\frac{2}{3}}\delta\Phi_b+\frac{2}{\sqrt3}\Phi_c\right]\nonumber\\
&+\cos \left[\frac{2\pi}{3}   \left(\tilde{N}_{gB}-\tilde{N}_{gC}\right)+\frac{\delta\Phi _a}{\sqrt{2}}+\frac{\delta\Phi _b}{\sqrt{6}}+\frac{2 }{\sqrt{3}}\Phi _c\right]\nonumber\\
&+\cos \left[\frac{2\pi}{3}   \left(\tilde{N}_{gC}-\tilde{N}_{gA}\right)-\frac{\delta\Phi _a}{\sqrt{2}}+\frac{\delta\Phi _b}{\sqrt{6}}+\frac{2 }{\sqrt{3}}\Phi _c\right]\Bigg),\label{sm3}
\end{align}
where we have introduced the shorthand notation $\tilde{N}_{g A}=N_{gA}-l_1$, $\tilde{N}_{g B}=N_{gB}-l_2$ and $\tilde{N}_{g C}=N_{gC}-l_3$ .

Using the effective action~\eqref{sm2}, the fluctuating fields in Eq.~\eqref{sm3} can then be integrated out, finally yielding the effective backscattering Hamiltonian 
\begin{align}
H^{(1)}_{\rm eff}=-\frac{1}{2}D^{1/3}\sqrt{E_C(e^\gamma\mu_0)^{1/3}}\left[r_{\rm eff}e^{i\frac{2 }{\sqrt{3}}\Phi _c}+{\rm c.c.}\right],\label{sgg1}
\end{align}
where we defined the effective reflection coefficient as 
\begin{align}
    r_{\rm eff}=r_i\left(e^{i\theta_{AB}}+e^{i\theta_{BC}}+e^{i\theta_{CA}}\right),\quad \theta_{ij}=2\pi\left(\tilde{N}_{gi}-\tilde{N}_{gj}\right)/3.\label{rud2}
\end{align}
Equations~\eqref{sgg1} and~\eqref{rud2} correspond, respectively, to the terms $\delta H_{ABC}$ and $\mathcal{R}_{\rm ABC}$ given in Eqs.~\eqref{eq:qb_ABC_reduction} and~\eqref{effectivec1} of the main text.
\subsection{Quantum critical points}
For a particular set of gate voltages $N_{gj}$, the effective reflection coefficient presented in equation~\eqref{rud2} vanishes, $r_{\mathrm{eff}}=0$. Then the destructive interference between backscattering processes completely suppresses backscattering events. In this case, the effective Hamiltonian describing our system becomes purely quadratic and can be solved exactly. These points in gate-voltage space are called critical points, and in the following we determine their precise locations.

Imposing the constraint for global minimum, $N_\Sigma=L$, and hence $\sum_j\tilde{N}_{gj}=0$, the effective reflection coefficient presented above can be written as $ r_{\rm eff}=r_i e^{\frac{2\pi i}{3}\left(\tilde{N}_{gA}-\tilde{N}_{gB}\right)}\left[ 1+ e^{2\pi i \tilde{N}_{gB}}+e^{-2\pi i \tilde{N}_{gA}}\right]$. Therefore, for the considered range $0\leq N_{gj}\leq 1$, $r_{\mathrm{eff}}$ vanishes precisely at two points $N_{gA}=N_{gB}=1/3$ and $N_{gA}=N_{gB}=2/3$. The corresponding $N_{gC}$ is determined by the constraint $N_\Sigma=L$, which finally yields the two critical points $(N_{gA},N_{gB},N_{gC})=(1/3,1/3,1/3)$ and $(N_{gA},N_{gB},N_{gC})=(2/3,2/3,2/3)$.
These two critical points are related to each other by a particle-hole transformation.

\subsection{Instanton tunneling}
From the previous subsection, the pinning energy of the outer potential is $V_{\mathrm{o,min}}=-3r_0\sqrt{D E_C}$, corresponding to the stationary points $\Phi_b=\frac{\pi}{2\sqrt6}(N_{gB}-N_{gA})+\frac{2\pi}{\sqrt6}(l_1-l_2)$ and $\Phi_a=\frac{\pi}{6\sqrt2}(N_{gA}+N_{gB}-2N_{gC})+\frac{2\pi}{3\sqrt2}(2l_3-l_1-l_2)$, which satisfy the minimization constraint $\sum_j N_{gj}=\sum_j l_j$. To evaluate the dimension of a possible relevant instanton, we take the initial and final fields to be $\Phi^{(i)}_b=\frac{\pi}{2\sqrt6}(N_{gB}-N_{gA})+\frac{2\pi}{\sqrt6}(l_1-l_2)$, $\Phi^{(i)}_a=\frac{\pi}{6\sqrt2}(N_{gA}+N_{gB}-2N_{gC})+\frac{2\pi}{3\sqrt2}(2l_3-l_1-l_2)$ and $\Phi^{(f)}_b=\frac{\pi}{2\sqrt6}(N_{gB}-N_{gA})+\frac{2\pi}{\sqrt6}(l'_1-l'_2)$, $\Phi^{(f)}_a=\frac{\pi}{6\sqrt2}(N_{gA}+N_{gB}-2N_{gC})+\frac{2\pi}{3\sqrt2}(2l'_3-l'_1-l'_2)$. Therefore, the field differences are $\Delta\Phi_a=\frac{2\pi}{3\sqrt2}\bigl[2(l'_3-l_3)-(l'_1-l_1)-(l'_2-l_2)\bigr]=\frac{2\pi}{3\sqrt2}(2d_3-d_1-d_2)$ and $\Delta\Phi_b=\frac{2\pi}{\sqrt6}(d_1-d_2)$. Because of the constraint $\sum_j N_{gj}=\sum_j l_j$, we always have $\sum_j d_j=0$. To find the smallest possible instanton dimension, we thus choose $(d_1,d_2,d_3)=(1,-1,0)$, which gives the instanton dimension $\Delta_{\rm inst}=\frac{|\Delta\Phi_a|^2+|\Delta\Phi_b|^2}{\pi^2}=\frac{8}{3}>1$, which is irrelevant.

\section{Strongly backscattering two inner links: AB criticality}
Here, we detail the physics in the regime in which the two inner QPCs exhibit strong backscattering relative to the remaining QPCs. For concreteness, we take QPCs $4$ and $5$ to be the inner QPCs with strong backscattering and assume $|\tilde{r}_4|=|\tilde{r}_5|=r_s$. Similarly, for the remaining QPCs, we assume for simplicity that $|\tilde{r}_1|=|\tilde{r}_2|=|\tilde{r}_3|=|\tilde{r}_6|=r_w$. The backscattering Hamiltonian~\eqref{sm1} can then be written as $H'_{\mathrm{b}}=H'_{\mathrm{bw}}+H'_{\mathrm{bs}}$ with
\begin{align}
H'_{\rm bw}=-r_w\sqrt{D E_C}&\Bigg(\cos \left[\frac{\pi}{2}  \left(2 N_{gA}+N_{gB}+N_{gC}\right)+\frac{\Phi _5}{\sqrt{6}}-\frac{\Phi _6}{\sqrt{2}}-\frac{2 \Phi _4}{\sqrt{3}}\right]\nonumber\\
&+\cos \left[\frac{\pi}{2}   \left(N_{gA}+2 N_{gB}+N_{gC}\right)-\sqrt{\frac{3}{2}} \Phi _5+\frac{\Phi _6}{\sqrt{2}}\right]\nonumber\\
&+\cos \left[\frac{\pi}{2}   \left(N_{gA}+N_{gB}+2 N_{gC}\right)+\frac{2 \Phi _4}{\sqrt{3}}+\sqrt{\frac{2}{3}} \Phi _5\right]\nonumber\\
&+\cos \left[\frac{\pi}{2}   \left(N_{gA}-N_{gC}\right)+\frac{2 \Phi _4}{\sqrt{3}}-\frac{\Phi _6}{\sqrt{2}}-\frac{\Phi _5}{\sqrt{6}}\right]\Bigg),
\end{align}
and
\begin{align}
H'_{\rm bs}=-r_s\sqrt{D E_C}&\Bigg(\cos \left[\frac{\pi}{2}   \left(N_{gA}-N_{gB}\right)+\sqrt{2} \Phi _6\right]\nonumber\\
&+\cos \left[\frac{\pi}{2}   \left(N_{gB}-N_{gC}\right)+\sqrt{\frac{3}{2}} \Phi _5+\frac{\Phi _6}{\sqrt{2}}\right]\Bigg).
\end{align}
As in the discussion in the previous section, the stationary points of $H'_{\mathrm{bs}}$ are $\theta'_1=2\pi n_1$, $\theta'_2=2\pi n_2$, with $n_j\in\mathbb{Z}$, where $\theta'_j$ denote the angles of the first and second cosine terms, respectively, in $H'_{\mathrm{bs}}$. The fixed point solutions are then given by $\Phi_6^*=\sqrt2 \pi n_1-\frac{\pi}{2\sqrt2}\left(N_{gA}-N_{gB}\right)$, and $\Phi^*_5=\sqrt{\frac{2}{3}}\left[2\pi n_2-\pi n_1+\frac{\pi}{4}\left(N_{gA}-3N_{gB}+2N_{gC}\right)\right] $. We then take the fluctuation around the stationary points as $\Phi_5=\Phi_5^*+\delta\Phi_5$, and $\Phi_6=\Phi_6^*+\delta\Phi_6$ to write the pinning potential in the form
$H'_{\rm bs} \sim H_{\rm const}+r^2_s E_C\frac{1}{4}\left[3\delta\Phi_5^2+2\sqrt3\delta\Phi_5\delta\Phi_6+5\delta\Phi_6^2\right]$. Then integrating out those massive fluctuation in $H'_{\rm bw}$, we finally arrive at the effective backscattering Hamiltonian
\begin{align}
H_{\rm eff}^{(2)}=-r_w 3^{1/24}\left(\frac{e^\gamma\mu_1}{D}\right)^{1/6}\sqrt{D E_C}&\Bigg(\cos \left[-\frac{2\pi}{3}  \left(2 N_{gA}+N_{gC}\right)+\frac{2 \Phi _4}{\sqrt{3}}\right]+\cos \left[\frac{2\pi}{3}   \left(N_{gA}+2 N_{gC}\right))+\frac{2 \Phi _4}{\sqrt{3}}\right]\nonumber\\
&\qquad\qquad+3^{1/8}\cos \left[\frac{2\pi}{3}   \left(N_{gA}-N_{gC}\right)+\frac{2 \Phi _4}{\sqrt{3}}\right]\Bigg),\label{rudra}
\end{align}
where $\mu_1 \sim r_s^2 E_C$, and we have retained only the terms relevant in the RG sense. The corresponding effective reflection coefficient in equation~\eqref{rudra} is presented as equation~\eqref{eq:qb_AB_amplitude} in the main text. By repeating the procedure outlined in the previous section, one can also show that the instanton tunneling events are completely irrelevant, with a lowest scaling dimension of $8/3$.

\begin{thebibliography}{87}
\bibitem [{ChargeProjection()}]{Jnote}%
  \BibitemOpen
  ChargeProjection,\ \href@noop {} {}\bibinfo {note} {To leading order in the
  charge-manifold projection, the effective tunneling amplitudes are
  $(J_1,J_2,J_3)=(t_4,t_5,t_6)$ for the isolated triangle and $(t_1,t_4,t_2^*)$
  for DCK. Virtual transitions to higher charge states renormalize effective
  couplings and triplet energies.}\BibitemShut {Stop}%
\end{thebibliography}
\end{document}